\documentclass[prl,aps,twocolumn]{revtex4}
\usepackage{graphicx}
\usepackage{dcolumn}
\usepackage{amsmath}

\begin{document}

\title{Coulomb Correlations and  Magnetic Anisotropy
in ordered $L1_0$ CoPt and FePt alloys}

\author{Alexander B. Shick \\
Institute of Physics ASCR, Na Slovance 2, 182 21 Prague 8, Czech
Republic  \\
Oleg N. Mryasov \\
Seagate Research, 1251 WaterFront Place,
Pittsburgh, PA, 15222, USA}


\begin{abstract}
We present results of the magneto-crystalline anisotropy energy (MAE) calculations for chemically ordered
$L1_0$  CoPt and FePt alloys
taking into account the effects of strong electronic
correlations and spin-orbit coupling.
The local spin density + Hubbard U approximation  (LSDA+U) is
shown to provide a consistent picture of the magnetic ground state properties
when intra-atomic Coulomb correlations are included for both  3$d$ and 5$d$ elements.
Our results  demonstrate significant and complex contribution of correlation effects
to large MAE of these material.
\end{abstract}


\maketitle

Recent progress in fabrication and characterization of the magnetic
nano-particles and thin films based on the
ordered $L1_{0}$  CoPt and FePt alloys
renewed interest in understanding the mechanisms contributing to the large
magneto-crystalline anisotropy of these materials.
While the earlier studies were primarily
motivated by permanent magnet  applications,
current research efforts are focused on
the use of CoPt and FePt alloys for the high density magnetic recording.
The large and controllable magnetic anisotropy energy (MAE) is then  a
crucial property
to overcome a superparamagnetic limit \cite{weller_para}.

The  chemically ordered $L1_{0}$ phase of FePt has large uniaxial MAE
($\approx 6.6*10^7 erg/cc$) which is almost by two orders of magnitude higher
than their disordered fcc phase (a dramatic MAE increase with chemical ordering
is also typical for other alloys of this kind: CoPt and FePd).
The cubic symmetry is broken in $L1_{0}$ phase by (i) stacking
of alternate planes of 3$d$ element (Fe or Co) and Pt along [001] direction; (ii) by tetragonal
distortion due to 3$d$ and 5$d$ atomic size mismatch. For the tetragonal crystal with
uniaxial symmetry, the MAE depends on polar $\theta$
magnetization angle  \cite{chikazumi} as,
\begin{eqnarray}
\label{eq:mae}
E = K_1 \sin^2{\theta} \; + \; K_2 \sin^4{\theta} + \mbox{...}
\end{eqnarray}
where $K_1$  and $K_2$ are the anisotropy constants.
The series Eq.(\ref{eq:mae}) is rapidly converging and
$K_1$ $>>$ $K_2$ for CoPt and FePt $L1_{0}$ alloys \cite{ermakov,ivanov}. The
MAE is then computed as the total energy difference
when magnetization
is oriented along [110] and [001] crystal axes (MAE = $E[110] \; - \; E[001]$).

A significant amount of work has been done to calculate this
energy difference from first principles, employing the local spin
density approximation (LSDA). For the case of CoPt alloy, the
fully relativistic Korringa-Kohn-Rostoker (KKR) method yields the
MAE of 0.058 meV/f.u. which is very different from results of
linear muffin-tin orbitals method (LMTO) calculations in
atomic-sphere-approximation (ASA):  2.29 meV \cite{sov}, 1.5 meV
\cite{sakuma}, 2 meV \cite{daalderop}, the
augmented-spherical-wave (ASW) result of 0.97 meV \cite{oppenier},
and full-potential LMTO (FLMTO) calculation results of 1.05 meV
\cite{ravindran} and 2.2 meV \cite{galanakis}. As can be seen,
there is an apparent and significant scatter of the theoretical
results, while the experimental measurements are known to be quite
accurate and consistent yielding the MAE of 1 meV \cite{ermakov}.
The situation is similar for the case of FePt alloy: the LMTO-ASA
calculations yield 3.4 meV \cite{sov}, 2.8 meV \cite{sakuma}, 3.5
meV \cite{daalderop}, ASW yields \cite{oppenier}  2.75 meV, and
FLMTO 3.9 meV \cite{galanakis} and 2.7 meV \cite{ravindran}, and
none of these calculations reproduce the experimental MAE of 1.3
$-$ 1.4 meV for the bulk \cite{ivanov}, and 1 $-$ 1.2 meV for the
films \cite{film} (when extrapolated to T=0 \cite{zeroext}).

The accurate {\it ab-initio} calculation of the MAE in itinerant
ferromagnet is a very difficult task due to its notorious sensitivity to
numerical details \cite{Trygg}.  More importantly,
the LSDA (or generalized gradient approximation (GGA))
which is conventionally used in the first-principles theory, lacks
proper orbital polarization due to Coulomb correlation effects \cite{igor}.
In this paper, we wish  (i) to clarify the ability of the LSDA theory to reproduce
the experimental MAE for CoPt and FePt alloys, when highly accurate full-potential
relativistic linearized augmented plane wave method (FP-LAPW) is used for the
total energy MAE calculations; (ii) to go beyond LSDA and to
investigate the role of electron correlations. We account for the on-site Coulomb
correlation effects by using the LSDA+U approach \cite{LDAU} and to show that correlations should be
included for both ``magnetic" (Fe and Co) and ``non-magnetic" (Pt) sites to describe
consistently the magnetic ground state properties, such as the MAE and spin and orbital
magnetic moments. For the first time, we quantitatively demonstrate a significant and
complex character of the intra-atomic Coulomb repulsion contribution to the MAE of
itinerant ferromagnet with strongly magnetic 3$d$ and non-magnetic 5$d$-elements.

We start with the conventional LSDA band theoretical method
together with the relativistic FP-LAPW method \cite{rlapw} and
apply them to perform total energy electronic and magnetic
structure calculations for the magnetization fixed along each of
[110] and [001] axes, respectively and the MAE. The special
$k$-points method is used for Brillouin Zone integration with the
Gaussian smearing of 1 mRy for $k$-points weighting \cite{Trygg}.
For convergence of the total energy differences within desired
accuracy (better than a few $\mu eV$), about 11000 $k$-points are
used (the MAE  as a function of $k$-points number is calculated to
be very similar to the results of Ref. \cite{ravindran}). The
experimental values for the lattice parameters ($a$  =  7.19 a.u.,
$c$ = 7.01 a.u. for CoPt, and $a$  =  7.30 a.u., $c$ = 7.15 a.u.
for FePt) are used \cite{ravindran}.
\begin{table}[t]
\caption{Magnetic Anisotropy Energy (meV/f.u.) calculated using FP-LAPW method
within the LSDA theory.
}
\begin{tabular}{ccccccccc}
\hline
                     &        & Ref. \cite{oppenier} & Ref. \cite{galanakis} & Ref. \cite{ravindran} &  Present\\
\hline
\multicolumn{2}{c}{\bf CoPt} & 0.97                      & 2.2                        & 1.05                 & 1.03\\
\multicolumn{2}{c}{Exp.}& \multicolumn{4}{c}{0.82 ($T=293$ K) \cite{ermakov}; 1.00 ($T=0$ K) \cite{zeroext}} \\
\hline
\multicolumn{2}{c}{\bf FePt} & 2.75                       & 3.9                        & 2.73                 & 2.68\\
\multicolumn{2}{c}{Exp.}& \multicolumn{5}{c}{0.7 $-$ 1.2 ($T=293$ K) \cite{ivanov,film}; 1 $-$ 1.4 ($T=0$ K) \cite{zeroext}} \\
\hline
\end{tabular}
\end{table}
The calculated MAE is shown in Table I. in comparison with  recent total energy
calculations and experimental results.
The present FP-LAPW results are
in very good quantitative agreement with ASW results of Oppeneer \cite{oppenier}
and FLMTO results of Ravindran \cite{ravindran} and disagree substantially
with FLMTO results of Galanakis \cite{galanakis} for the reason which is unclear to us.
For CoPt alloy the MAE is calculated in a very good agreement with experimental data, while
for FePt our LSDA results (together with those of Refs. \cite{oppenier,ravindran}) overestimate
value of MAE by a factor of two.

Very recently, the electron-electron interaction was shown to play an important role for the MAE in itinerant
$d$ \cite{savrasov} and $f$-electron \cite{sasha} magnetic materials.
Here, we use the LSDA+U method combined with relativistic FP-LAPW basis \cite{sasha,SLP}
to account for the intra-site Coulomb repulsion $U$. Minimization of the LSDA+U total energy
functional with SOC treated self-consistently \cite{rlapw} generates not only the ground state
energy but also one-electron energies and states providing the orbital contribution to the
magnetic moment. The basic difference of LSDA+U calculations from the LSDA is its explicit
dependence on on-site spin and orbitally resolved occupation matrices. The LSDA+U creates
in addition to spin-only dependent LSDA potential, the spin and orbitally dependent
on-site ``$+U$" potential which produces the orbital polarization \cite{scf}.
Since the LSDA+U method is rarely applied to metals,
the appropriate values of intra-atomic repulsion $U$ for Fe, Co and Pt atoms in metals
are not known precisely as they are strongly affected by the metallic screening.
Here we choose $U_{3d}$ from the range of 1-2 eV,
and $U_{5d}$ from the range of 0-1 eV (according to
the experimental values for pure
metals \cite{steiner}) in order to reproduce experimental values of the MAE.
For the exchange
$J$ we use the values of $J_{{Co}} = 0.911 \; \mbox{eV}, \; J_{{Fe}} = 0.844 \; \mbox{eV},
\; J_{{Pt}} = 0.544 \; \mbox{eV}$ which are obtained as a result of constrained LSDA
calculations \cite{SDA}, and are close to their atomic values.
\begin{figure}[h]
\centering
\includegraphics[width=6.5cm,clip]{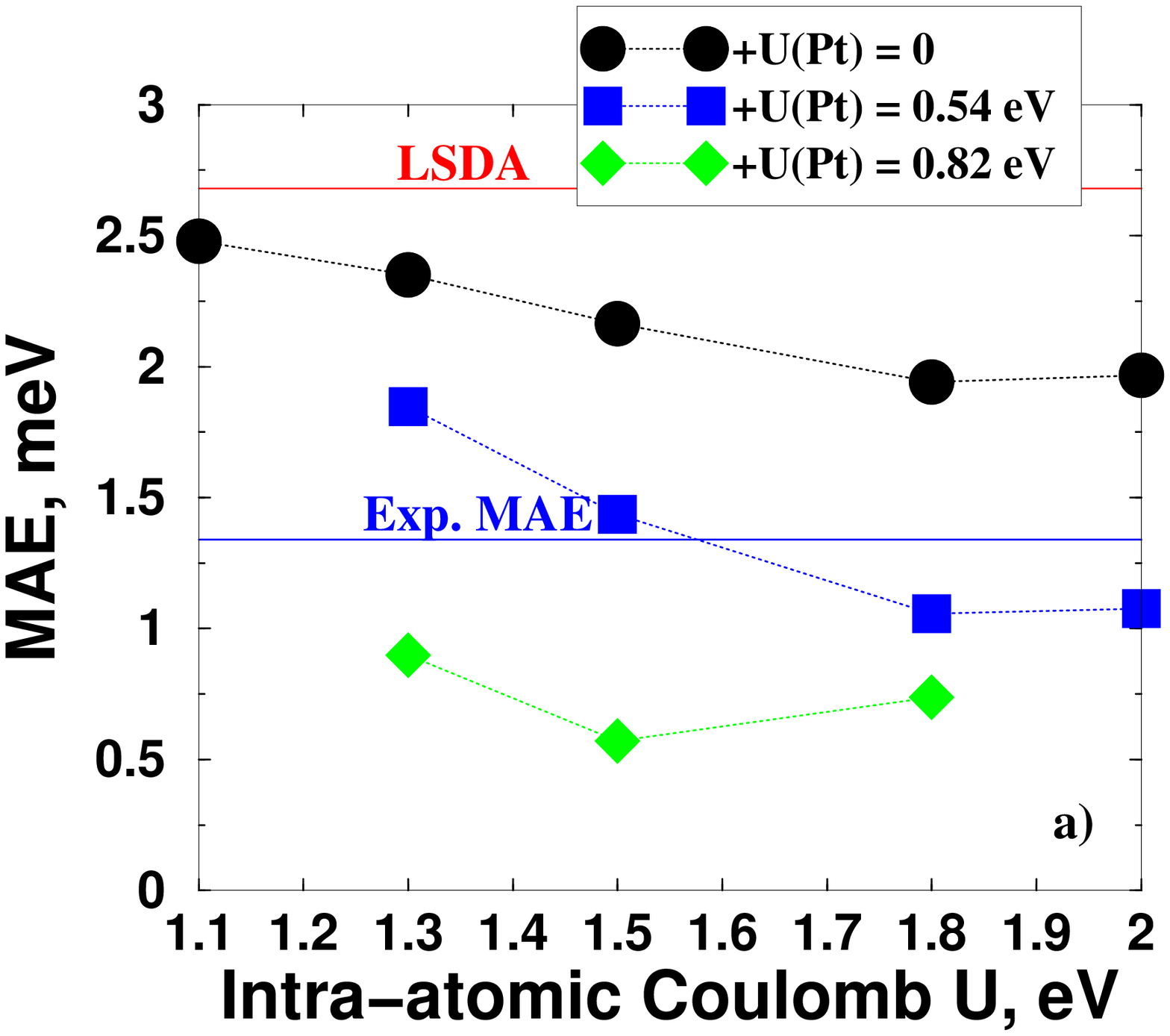}
\includegraphics[width=6.5cm,clip]{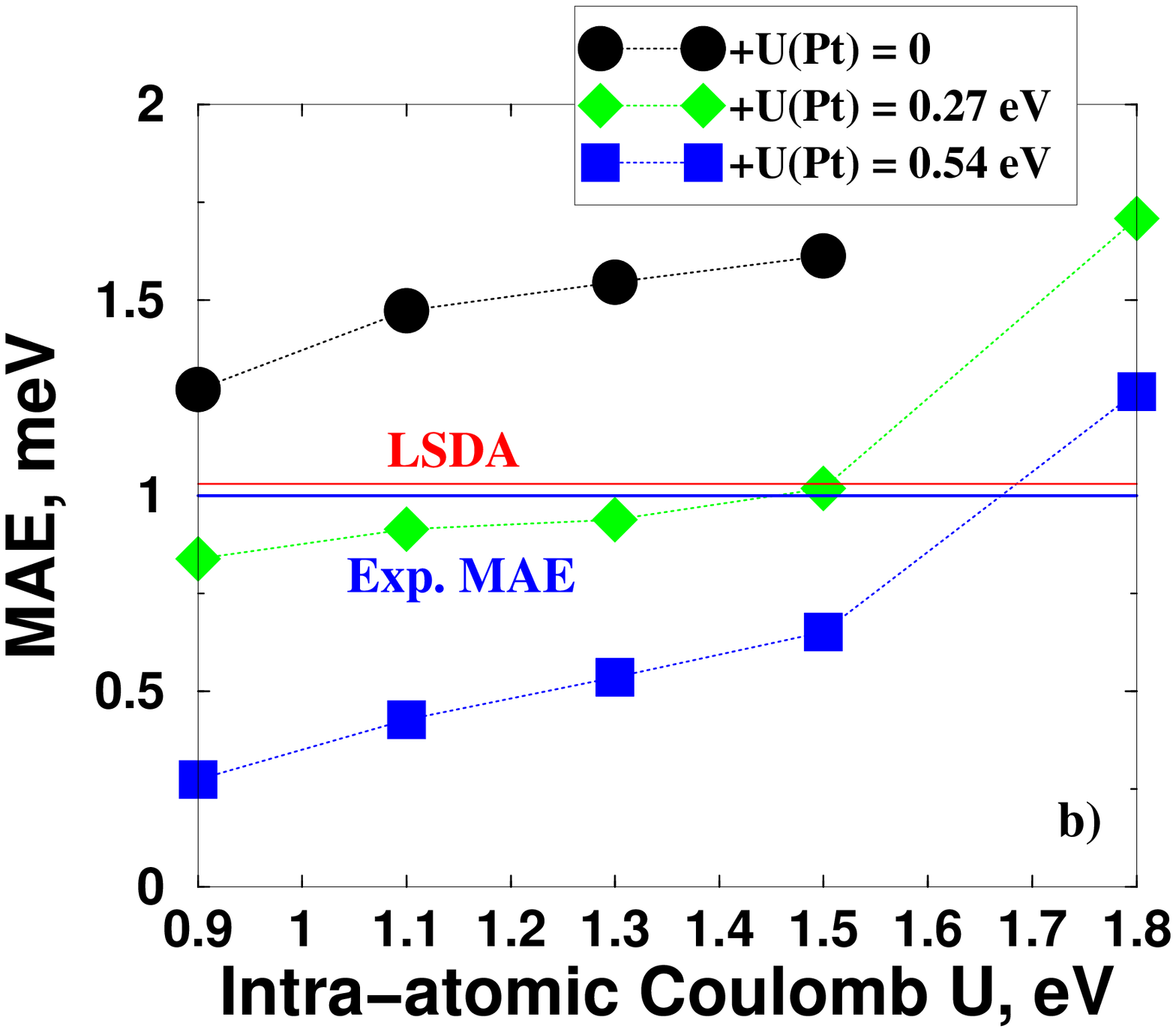}
\caption{(color) The MAE vs intra-atomic Coulomb repulsion ($U$)
on 3$d$-site for (a) FePt alloy; (b) CoPt alloy for different
values of $U$ on Pt-site. Note, that we use the bulk experimental
MAE values \cite{ermakov,ivanov} extrapolated to $T=0$ K.
}
\end{figure}

First, we discuss the FePt alloy. The MAE as a function of $U_{{Fe}}$ is shown
in Fig. 1(a). When $U$ is included on Fe-site only and varied in the interval of 1-2 eV,
the MAE is decreasing from its LSDA value. The experimental
value of the MAE can not be reached without unreasonable increase of $U$.
Further meaningful reduction of the  MAE can be achieved by
including $U$ on the Pt-site (see, Fig. 1(a)) and for the value of
$U_{{Pt}}$ around 0.544 eV ($=J_{{Pt}}$) the MAE  observed for the
bulk FePt can be reproduced. We note that  the correct value of
the MAE can be obtained in the region of $U_{{Fe}}$ $\approx$
1.5-1.6 eV, $U_{{Pt}}$ $\approx$ 0.5-0.6 eV.

We choose the  \{$U_{{Fe}}$ = 1.516 eV, $U_{{Pt}}$ = 0.544 eV\} as
a representative material specific parameters set to analyze the
spin and orbital ground state properties. The calculated spin
$M_s$, orbital $M_l$ and total magnetic moments for the
magnetization directed along [001] axis are shown in Table II.
The LSDA+U  agrees with observed total magnetization/f.u. $M^{tot}$, yields a slight
increase of $M_s^{{Fe}}$ from its LSDA value of 2.88 $\mu_B$ and substantially increases the
$M_l^{{Fe}}$ from 0.0657 $\mu_B$. The change of $M_s$ and $M_l$ of Pt-site is very small
as compared with LSDA.  The effect of $U$ on the
$M_l$ is seen to be similar to the ``orbital polarization correction"
of Brooks {\it et al.} (LSDA+OP) \cite{ravindran} (see, Table II). This {\it ad-hoc} LSDA+OP
correction is known to improve the $M_l$ in a number of $d$- and
$f$-electron compounds, but it does not reproduce the MAE for FePt alloy.
\begin{table}[t]
\caption{Spin ($M_s$), Orbital ($M_l$) magnetic moments for
3$d$ and 5$d$ atoms, and Total Magnetic Moment ($M^{tot}$) per formula unit ($\mu_B$);
Magnetic Anisotropy Energy (meV/f.u.) as results of LSDA+U calculations
for FePt
and CoPt
alloys. The LSDA+U calculated $M_s$ and $M_l$ values for bcc-Fe and hcp-Co obtained
with the same values for $U_{3d}$ and $J_{3d}$ as for Fe- and Co-atoms in FePt and
CoPt alloys.}
\begin{tabular}{ccccccccc}
\hline
\multicolumn{2}{c}{\bf FePt[001]} & MAE  & $M^{tot}$ & Atom  &  $M_s$ & $M_l$ \\
\hline
\multicolumn{2}{c}{LSDA+U} & 1.3   &  3.47          & Fe    &  3.00  & 0.114  \\
\multicolumn{4}{l}{\{$U_{{Fe}}$ = 1.52 eV, $U_{{Pt}}$ = 0.54 eV\}}& Pt    &  0.34  & 0.048  \\
\hline
\multicolumn{2}{c}{LSDA+OP}& 2.9   &  3.36          & Fe    &  2.89  & 0.110  \\
             &             &       &                & Pt    &  0.35  & 0.048  \\
\hline
\multicolumn{2}{c}{Exp.}   & 1.3   &  3.4           &       &        &        \\
\hline
\multicolumn{2}{c}{\bf CoPt[001]} & MAE  & $M^{tot}$ & Atom  &  $M_s$ & $M_l$ \\
\hline
\multicolumn{2}{c}{LSDA+U} & 1.0   &  2.55          & Co    &  1.93  & 0.253 \\
\multicolumn{4}{l}{\{$U_{{Co}}$ = 1.70 eV, $U_{{Pt}}$ = 0.54 eV\}} & Pt    &  0.38  & 0.065 \\
\hline
\multicolumn{2}{c}{LSDA+OP}& 1.64  &  2.37          & Co    &  1.80  & 0.161 \\
             &             &       &                & Pt    &  0.39  & 0.062 \\
\hline
\multicolumn{2}{c}{Exp.}   & 1.0   &  2.4           & Co    & 1.9-2.1 &0.284 \\
             &             &       &                & Pt    &  0.36   &0.090 \\
\hline
\multicolumn{2}{c}{\bf bcc-Fe[001]}& & &                    &  $M_s$ & $M_l$ \\
\multicolumn{2}{c}{LSDA+U}         &\multicolumn{3}{c}{$U_{Fe}$ = 1.52 eV} & 2.234  & 0.085 \\
\multicolumn{2}{c}{LSDA+OP \cite{Trygg}}&        &   &            & 2.193  & 0.078 \\
\multicolumn{2}{c}{Exp.}&        &   &            & 2.13   & 0.08  \\
\hline
\multicolumn{2}{c}{\bf hcp-Co[0001]}&&&                                     & $M_s$ & $M_l$ \\
\multicolumn{2}{c}{LSDA+U}          &\multicolumn{3}{c}{$U_{Co}$ = 1.70 eV} & 1.631  & 0.153 \\
\multicolumn{2}{c}{LSDA+OP \cite{Trygg}}&        &   &             & 1.591  & 0.123 \\
\multicolumn{2}{c}{Exp.}&        &   &             & 1.52   & 0.14  \\
\hline
\end{tabular}
\end{table}

We now describe CoPt alloy.  The MAE as a function of $U_{{Co}}$ is shown
in Fig. 1(b). When $U$ is included on Co-site only and varied in the interval of 1-2 eV,
the MAE increases from its LSDA value and deviates from the experiment. As in the case
of FePt alloy, we need to include $U$ on the Pt-site. The increase of $U_{{Pt}}$ yields the
decrease of the MAE and its correct value is found in the region of
\{$U_{{Co}}$ $\approx$ 1.4-1.7 eV, $U_{{Pt}}$ $\approx$ 0.3-0.6 eV\}.
We choose the set \{$U_{{Co}}$ = 1.698 eV, $U_{{Pt}}$ = 0.544 eV\} as a representative
(note, that we use the same value of $U_{{Pt}}$ as for the FePt alloy, indicating its transferability).
The calculated ground state $M_s$, $M_l$ and $M^{tot}$
for the magnetization directed along [001] axis are shown in Table II.
The $M^{tot}$ is calculated
in a good quantitative agreement with results of LSDA+OP and experimental data \cite{daalderop}.
There is only small increase in $M_s^{{Co}}$ from LSDA value of 1.81 $\mu_B$ while the
$M_l^{{Co}}$ is substantially enhanced from its LSDA value of 0.093 $\mu_B$ (Table II).
The change of $M_s$ and $M_l$ for Pt-site is very small. The $M_l^{{Co}}$ increase
in LSDA+U calculations is substantially more pronounced than obtained in LSDA+OP approximation \cite{ravindran}.

Recently, the element-specific $M_s$ and $M_l$ for CoPt were measured by Grange {\it et al.}
using the X-ray Magnetic Circular Dichroism (XMCD) \cite{grange}.
The LSDA+U calculated values for $M_s$ and $M_l$ for the Co-site are in a good quantitative agreement
with XMCD results \cite{XMCD}. For the Pt-site,
the agreement is not as good.
Probably it is caused by the use of the atomic-like sum rules to extract
$M_s$ and $M_l$ from XMCD spectra. This procedure is not reliable for Pt
due to substantially itinerant character of Pt 5d-electrons.
The LSDA+U calculations reproduce consistently
the MAE, total, spin and orbital ground state magnetic moments for CoPt alloy (Table II).  Both LSDA and LSDA+OP
are only partially successful: LSDA yields correct value of MAE but fails for $M_l^{Co}$ and LSDA+OP
improves somewhat $M_l$ but does not reproduce the MAE.

To evaluate further  the consistency of the LDSA+U results, we
performed the LSDA+U calculations for the elemental
$3d$-ferromagnet bcc-Fe and hcp-Co with the same values of the
$U_{3d}$ and $J_{3d}$ as for Fe- and Co-atoms in FePt and CoPt
alloys (see, Table II). It is seen that without any further
adjustments of parameters the LSDA+U provides very reasonable results for the
orbital magnetization in elemental $3d$ ferromagnet.
These results demonstrate  that the on-site Coulomb interaction
parameters $U_{3d}$ are well transferable in the transitional
$d$-metal systems.

As for the choice of the Coulomb-$U$ for Pt, the challenge lies in
correcting the LSDA orbital polarization without harming the
exchange splitting which is expected to be well accounted in  the
LSDA. The choice of  $U_{Pt}=$0.5-0.6 eV looks then quite
reasonable. Indeed, for the  Pt-5$d$ states having almost equal
on-site occupations, the choice of $U_{Pt} \approx J_{Pt}$
corresponds to an effective Stoner exchange $I_{LSDA+U} \approx
I_{LSDA}$ \cite{Mazin} preserving the LSDA spin polarization. This
allows to ensure that  the LSDA+U correction contributes entirely
to the orbital polarization. The LSDA+U method, while proposed to
deal with the problems specific for the localized states, in fact
is not limited by this case. This method can be used as soon as
on-site Coulomb correlation in the form of the Hubbard model is
physically meaningful. The above comparison with  available
experimental data for the MAE and spin/orbital magnetic moments,
and physically reasonable choice of parameters justify the use of
on-site  LSDA+U correction \cite{LDAU} for the Pt-5$d$ states as
the way to correct on-site orbital polarization.

We now discuss the relation between the MAE and the anisotropy of
the orbital magnetic moment $\Delta M_l (= M_l \parallel [110] -
M_l \parallel [001])$. Bruno \cite{PB} showed that in the limit of
strong exchange splitting $\Delta_{ex}$ $>>$ SOC, the MAE is
proportional to $\Delta M_l$. This model predicts the positive MAE
of 0.2 meV for CoPt and negative MAE of -2.1 meV for FePt, in
disagreement with the total energy calculations and experiment
(Table II).

A more general form for MAE was given in Ref. \cite{VDL}:
\begin{equation}
\label{eq:vdl}
\mbox{MAE} \approx - \frac{\xi}{4} \Delta (M^{\downarrow}_l - M^{\uparrow}_l) + \Delta E_T [{\uparrow \downarrow}
\mbox{-``spin-flip"}]
\end{equation}
where $\xi$ is the SOC constant.
The 1st term ($\Delta E_L$) is the
$\uparrow \uparrow; \downarrow \downarrow$-spins contribution
due to the orbital moment $\vec{L}$, and the 2nd term
couples the  $\uparrow \downarrow$-spins and is related to the spin magnetic dipole moment $\vec{T}$.
In the limit of SOC $<<$ $\Delta_{ex}$,
the $\Delta E_T \approx - {3 {\xi}^2/}{\Delta_{ex}}[\Delta Q_{zz}]$ is proportional to the
difference of quadrupole moments $Q_{zz}$ for $z=[110], \; [001]$.
Note, that this $\Delta E_T$ form is valid for Fe and Co since their
SOC (0.07-0.08 eV) $<<$ $\Delta_{ex}$ (3-4 eV) and can not be used for Pt which has
the SOC (0.6 eV) $>$  $\Delta_{ex}$ (0.2 eV) \cite{SDKW}.

The Eq.(\ref{eq:vdl}) gives  for the CoPt alloy $\Delta E_L^{Co}$=1.4
meV, $\Delta E_T^{Co}$=-0.7 meV and $\Delta E_L^{Pt}$=1.9 meV, and
for FePt alloy, $\Delta E_L^{Fe}$=-0.1 meV, $\Delta E_T^{Fe}$=1.0
meV, and $\Delta E_L^{Pt}$=1.2 meV. Here,  large $\Delta
E_T^{Fe,Co}$ contributions to the MAE naturally originate from the
difference in the inter-plane $\{xz,yz\} \; 3d-5d$ and in-plane
$\{xy\} \; 3d-3d$ hybridization. Without $\Delta E_T^{Pt}$
contribution, the total MAE of 2.7 meV (CoPt) and 2.1 meV (FePt)
can be estimated, exceeding substantially the experimental values
(cf., Table II). We can only roughly estimate that $\Delta E_T^{Pt}  \sim
- \Delta Q_{zz} (= \mbox{0.12 (CoPt),0.09 (FePt)}$ provides
additional negative MAE contributions which are expected to reduce
a total MAE  towards the experimental data. Thus, due to the
strong Pt-SOC, neither of commonly used MAE parameterizations
\cite{PB,VDL} based on SOC-perturbation theory expansions is valid
on the quantitative and do not provide a substitute for the total
energy MAE calculations.

Still, it is of  interest  to apply the Eq. (\ref{eq:vdl}) to
analyze qualitatively the origin of
 MAE dependence on $U$ shown in Fig.1.
This analysis shows that LSDA+U MAE vs $U$ dependence is
qualitatively consistent with the Eq.(\ref{eq:vdl}),
and the change in $\Delta E_L^{Pt} \sim \Delta (M^{\downarrow}_l - M^{\uparrow}_l)^{Pt}$
contributes substantially to the MAE variations with the $U$.
In particular, we find  the ``coupling" between $U_{3d}$ and
$\Delta E_L^{Pt}$ which originates from strong $3d-5d$ hybridization,
so that tiny $U_{3d}$-induced changes in Pt-$\Delta (M^{\downarrow}_l - M^{\uparrow}_l)$
produce substantial MAE change due to the strong Pt-SOC. It also explains
surprisingly strong MAE dependence on $U_{Pt}$ (cf., Fig.1), as
a variation of $U_{Pt}$ causes tiny Pt-$\Delta (M^{\downarrow}_l - M^{\uparrow}_l)$
change
($\sim 10^{-3}$ $\mu_B$) which in turn changes the MAE substantially
due to the strong Pt-SOC.

To summarize, accounting for on-site Coulomb correlations beyond
what is included in LSDA, and using the LSDA+U method in a very
general implementation including SOC we have provided a
microscopic theory of the ground state magnetic properties in
$L1_0$ FePt and CoPt alloys. It is shown by comparison with the
experiment that LSDA+U method is capable of describing
quantitatively the MAE and orbital magnetization in these alloys
with physically reasonable choice of Coulomb-$U$ parameters. Using
the SOC-perturbation theory model we provide  interpretation
of our  numerical results.
These results are believed to be important for quantitative microscopic
understanding of the large MAE in these material, and will assist
in the development of the next generation magnetic recording
devices operating above $Tbit/in^2$ recording densities.

We are grateful to D. Weller, V. Drchal and P. Oppeneer for
helpful comments and useful discussion.


\vspace{-0.5cm}
\begin{thebibliography}{99}
\bibitem{weller_para} D. Weller and A. Moser,  IEEE Trans. Magn., {\bf 36}, 10 (1999);
 S. Sun {\it et al.}, Science {\bf 287}, 1989 (2000).

\bibitem{chikazumi} S. Chikazumi, {\it Physics of Magnetism} (Krieger, Malabar, FL, 1986).

\bibitem{ermakov} A. Ye. Yermakov and V.V. Maykov, Phys. Met. Metall.
{\bf 69}, 198 (1990) and references therein.

\bibitem{ivanov} O. A. Ivanov {\it et al.}, Phys. Met. Metall. {\bf 35}, 92 (1973).


\bibitem{staunton} S. Razee {\it et al.}, Phys. Rev. Lett. {\bf 82}, 5369 (1999).

\bibitem{sov} I. V. Solovyev {\it et al.}, Phys. Rev. {\bf B52}, 13419 (1995).

\bibitem{sakuma} A. Sakuma, J. Phys. Soc. Jpn. {\bf 63}, 3053 (1994).

\bibitem{daalderop} G.H.O. Daalderop {\it et al.}, Phys. Rev. {\bf B44}, 12054 (1991).

\bibitem{oppenier} P. Oppeneer,  J. Magn. Magn. Mater. {\bf 188}, 275 (1998).

\bibitem{ravindran} P. Ravindran {\it et al.}, Phys. Rev. {\bf B63}, 144409 (2000).

\bibitem{galanakis} I. Galanakis, M. Alouani and H. Dreysee,  Phys. Rev. {\bf B62}, 6475 (2000).

\bibitem{film}
J.U. Thiele {\it et al.}, J. Appl. Phys. {\bf 84}, 5686 (1998);
J.U. Thiele {\it et al.}, {\it ibid.} {\bf 91}, 6595 (2002);
S. Okamoto {\it et al.}, Phys. Rev. {\bf B66}, 024413 (2002).

\bibitem{zeroext} We use E.R. Callen and H.B. Callen theory (J. Phys. Chem. Solids {\bf 27},
1271 (1966)) and experimental magnetization $M(T)$ dependence \cite{ermakov,ivanov} to
find MAE($T \rightarrow 0$).

\bibitem{Trygg} J. Trygg  {\it et al.}, Phys. Rev. Lett. {\bf 75}, 2871 (1995).
The numerical convergence of the calculated difference in the total energy was analysed following
this work.

\bibitem{igor} I. V. Solovyev, A. I. Liechtenstein, and K. Terakura, Phys. Rev. Lett. {\bf 80}, 5758  (1998).

\bibitem{LDAU} V.I. Anisimov, F. Aryasetiawan and A. I. Lichtenstein,
J. Phys.: Condens. Matter {\bf 9}, 767 (1997).

\bibitem{rlapw} A. B. Shick, D. L. Novikov, and A. J. Freeman, Phys. Rev. {\bf B56}, R14259 (1997).


\bibitem{savrasov} I. Yang, S. Savrasov and G. Kotliar, Phys. Rev. Lett. {\bf 87}, 216405 (2001).

\bibitem{sasha}A. B. Shick and W. E. Pickett, Phys. Rev. Lett. {\bf 86}, 300 (2001).

\bibitem{SLP} A. B. Shick, A. I. Liechtenstein, and W. E. Pickett,
Phys. Rev. {\bf B 60}, 10763 (1999).


\bibitem{scf}
In these calculations, both $V_{LSDA}$ and $V_{+U}$
are determined self-consistently. The charge/spin densities which are needed for the $V_{LSDA}$
are converged better than $5. \times 10^{-5}$ electron/(a.u.)$^3$ in order to achieve the
total energy convergence better than few $\mu$eV.

\bibitem{steiner} M.M. Steiner, R.C. Alberts and L.J. Sham, Phys. Rev. {\bf B 45}, 13272 (1992).

\bibitem{SDA} I.V. Solovyev, P. H. Dederichs and V.I. Anisimov, Phys. Rev. {\bf B 50}, 16861 (1994); Note
that
small variations of $J$ are found not to affect the results of calculations.

\bibitem{grange} W. Grange {\it et al.}, Phys. Rev. {\bf B62}, 1157 (2000).

\bibitem{XMCD} The XMCD measures the ratio of $M_s$ and $M_l$ to the number of holes
$n_d$ in d-shell. To compare with LSDA+U results, the calculated $n_d^{{Co}}$ =2.845 and
$n_d^{{Pt}}$ = 2.370 were used together with the experimental $M_{s,l}/n_d$ ratios \cite{grange}.

\bibitem{PB} P. Bruno, Phys. Rev. {\bf B 39}, 865 (1989).

\bibitem{VDL} G. van der Laan, J. Phys.: Cond. Matter {\bf 10}, 3239 (1998).


\bibitem{SDKW}
A. B. Shick {\it et al.},
Phys. Rev. {\bf B 54}, 1610 (1996).

\bibitem{Mazin} I. Mazin {\it et al.}, cond-mat/0206548 (2002).


\end{thebibliography}
\end{document}